\documentclass[twocolumn,showpacs,preprintnumbers,amsmath,amssymb,prb]{revtex4}
\usepackage{dcolumn}
\usepackage{bm}
\usepackage[dvips]{graphicx}
\usepackage{subfigure}
\usepackage{longtable}
\usepackage[colorlinks=true, linkcolor=blue, citecolor=blue, urlcolor=cyan]{hyperref}
\usepackage{dcolumn}

\begin{document}

\title{High pressure behavior of CsC$_{8}$ graphite intercalation compound}

\author{N. Rey$^1$}
\author{P. Toulemonde$^{1,2}$}
\author{D. Machon$^1$}
\author{L. Duclaux$^3$}
\author{S. Le Floch$^1$}
\author{V. Pischedda$^1$}
\author{J.P. Iti\'e$^4$}
\author{A.-M. Flank$^4$}
\author{P. Lagarde$^4$} 
\author{W. A. Chrichton$^5$}
\author{M. Mezouar$^5$}
\author{Th. Str\"assle$^6$}
\author{D. Sheptyakov$^6$}
\author{G. Montagnac$^7$}
\author{A. San-Miguel$^1$}
\email{Alfonso.San.Miguel@lpmcn.univ-lyon1.fr}

\affiliation{$^1$Laboratoire de Physique de la Mati\`ere Condens\'ee et Nanostructures, Universit\'e Lyon 1; CNRS, UMR 5586, F-69622 Villeurbanne, France}
\affiliation{$^2$Institut N\'eel - D\'epartement MCMF, CNRS and Universit\'e
Joseph Fourier, 25 avenue des Martyrs, BP 166, F-38042 Grenoble
cedex 9, France.
}%
\affiliation{$^3$Laboratoire de Chimie Moléculaire et Environnement, Universit\'e de Savoie-ESIGEC, Campus Scientifique de Savoie-Technolac, Le Bourget du Lac Cedex, France}

\affiliation{$^4$
Synchrotron-SOLEIL, CNRS-UR1, BP48, 91192 Gif-sur-Yvette Cedex France
}%

\affiliation{$^5$
European Synchrotron Radiation Facility, rue 6 Jules Horowitz, B.P. 220 38043 Grenoble Cedex 09, France}%

\affiliation{$^6$
Laboratory for Neutron Scattering, ETH Zurich \& Paul Scherrer Institut, CH-5232 Villigen PSI, Switzerland
}%

\affiliation{$^7$
Laboratoire de Sciences de la Terre, UMR 5570 CNRS-ENSL-UCBL, 46 All\'ee d'Italie, F-69364 Lyon Cedex 07, France}


\begin{abstract}
The high pressure phase diagram of CsC$_{8}$ graphite intercalated compound has been investigated at ambient temperature up to 32 GPa. Combining X-ray and neutron diffraction, Raman and X-ray absorption spectroscopies, we report for the first time that CsC$_8$, when pressurized, undergoes phase transitions around 2.0, 4.8 and 8~GPa. Possible candidate lattice structures and the transition mechanism involved are proposed. We show that the observed transitions involve structural re-arrangements in the Cs sub-network while the distance between the graphitic layers is continuously reduced at least up to 8.9~GPa. Around 8~GPa, important modifications of signatures of the electronic structure measured by Raman and X-ray absorption spectroscopies evidence the onset of a new transition.
\end{abstract}


\pacs{Valid PACS appear here}

\maketitle

\section{Introduction}

Graphite intercalation compounds (GICs), well-known basic two-dimensional systems, have been extensively studied as template layered materials, allowing the intercalation of numerous different chemical species ($\gg$~100)~\cite{Dresselhaus02}. A recent renew of interest on GICs has been motivated by the discovery of the highest superconducting critical temperature $T_c$ of this family of materials in CaC$_6$ (11.5 K) and YbC$_6$ (6.5 K)~\cite{Weller05,eme05} leading to subsequent experimental~\cite{kim06,lamura06,bergeal06,sutherland07,hinks07,cubitt07} and theoretical~\cite{calandra05,csanyi05,mazin05,calandra06,mazin06,sanna07} efforts. 

Pressure is a crucial thermodynamic parameter in the understanding of electronic and structural properties of these low dimensional guest-host systems. Pressure effects include the enhancement of the host-guest interactions. In addition, under high pressure (HP) and high temperature (HT), GICs can constitute precursors for the formation of new carbon intercalated materials such as carbon clathrates~\cite{STou05}, which could show promising superconductivity~\cite{connet03} and mechanical~\cite{blase04} properties. 

In GICs, several structures associated with the layered topology like staging, in-plane decoration or stacking are observed. One of the most remarkable structural features is the staging phenomenon occurring at long-range order~\cite{Dresselhaus02}. Staging refers to the one-dimensional periodic alternation of filled and empty graphitic galleries. Stage-$n$ is defined as the structure in which intercalates are accommodated regularly in every $n$-th graphite layers. Thus for stage-1, all the graphitic galleries are periodically occupied by the intercalate layers. The repeat distance $I_c$ is defined as the distance along the $c$-axis between two successive intercalated layers. Important structural features are distinguished according to the value of the stage index-$n$. Two different decorations of intercalated layers commensurate to the graphite lattice for the stage-1 compounds have been observed: an hexagonal 2$\times$2 superlattice (Rb, Cs and K) and an hexagonal $\sqrt{3}\times\sqrt{3}$ superlattice (Li, Ca, Yb and others). In higher stage compounds, the metal layer arrangement can lead to an incommensurate modulation with respect to the graphite layers or can be even similar to a \textit{liquid}-like behavior~\cite{parry77,rousseaux90}.

Among the donor compounds, the family of K-GICs has been used as model for the description of pressure-induced transitions. The data from different studies were limited to relatively low values of pressure ($<$3 GPa). In the stage-1 KC$_8$, staging transitions occurred with the presence of a stage-2 and a fractional stage-3/2 in the 0-3.0~GPa range~\cite{fuerst83}. Bloch et al observed a $\sqrt{3}\times\sqrt{3}$ superlattice in the same pressure range~\cite{bloch85}. These structural changes in the decoration and in the staging are accompanied with an anomaly in the resistivity at 1.5~GPa~\cite{fuerst81}. No staging transitions were found in CsC$_8$ and RbC$_8$ up to 1~GPa~\cite{wada81} although there were observed pressure-induced anomalies in the resistivity of these two compounds at 2.0~GPa. In the stage-1 CaC$_6$, pressure induces the enhancement of superconductivity and near 8~GPa a structural instability was found when the critical temperature T$_c$ suddenly drops~\cite{gauzzi07}. Similar behavior of T$_c$ in YbC$_6$ is observed at 1.8~GPa~\cite{smith06}.  

For "diluted" intercalation, i.e.\ $n\geq$2, staging transitions $n\rightarrow n$+1 or $n$+2  were observed for moderate pressures 0.5-1.0 GPa. First staging transition from a stage-2 to stage-3 was evidenced in KC$_{24}$ by Fuerst et al~\cite{fuerst80}. This work has been extended to higher stage K-GIC and to high-stage Rb and Cs GICs by Wada~\cite{wada81}.  

CsC$_8$ is a good candidate to investigate the size effect on the guest-host interaction with pressure and to give new insight into the high-pressure properties and stability of this important class of materials. The total lack of information for P$>$2~GPa prompted us to examine this compound at higher pressure. We have hereby explored the high-pressure evolution up to 32~GPa of CsC$_8$. Information on the Cs sub-network has been obtained using X-ray diffraction thanks to the high Z of the guest species. On the other side, neutron diffraction appears as more sensitive to the carbon sub-network. Finally, local probes that constitute Raman and X-ray absorption spectroscopies are an excellent complement of the long-range order technique that is diffraction. This multi-technique approach is a key point for the understanding of the high pressure behavior of this highly reactive and complex ordered material. In particular, we have investigated the pressure-induced effects such as charge transfer between the guest and host species, the structural modifications in the Cs and the graphitic sub-networks. 

\section{Experimental environment}

\subsection{Synthesis route and structural characterization}
CsC$_{8}$ powder samples were synthesized from a high purity stoichiometric mixture of powder graphite (Aldrich, $<$20 $\mu$m) and Cs metal melt (Alfa Aesar, 99.98\%) in an argon filled glove-box~\cite{los06}. A 48h static vacuum annealing of the sample sealed in an ampoule allowed to achieve the intercalation process. Brown/reddish typical color of the stage-1 heavy alkali GICs was observed after the treatment.  
\begin{figure}
\includegraphics[width=0.5\textwidth]{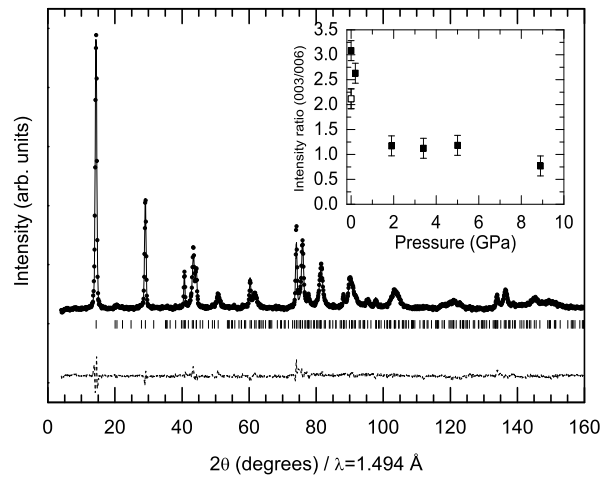}
\caption{\label{fig:npd_amb} Ambient pressure and temperature neutron powder diffraction pattern of CsC$_8$ fitted within a Rietveld refinement. The pattern was collected in a vanadium container. The inset shows the relative intensity of the (003) and (006) lines with pressure. Open symbol for decompression at zero pressure.} 
\end{figure}
Figure~\ref{fig:npd_amb} shows the neutron powder diffraction pattern of our CsC$_8$ powder measured at ambient conditions together with the residual of the Rietveld refinement using the \textsc{Fullprof} software \cite{rodriguez93}. Our structural model includes an anisotropic size broadening effect. A better quality refinement was found when the Cs atoms are in the \textit{3d} Wyckoff position ($\chi^2$=2.73) rather than in the \textit{3b} position ($\chi^2$=3.74) as Gu\'erard \textit{et al} suggested~\cite{guera78}. A \textit{3b} position would imply a $\alpha\alpha\alpha$ stacking sequence for the Cs atoms whereas the stacking sequence for this compound is supposed to be $\alpha\beta\gamma$ which can be reproduced with the \textit{3d} setting. The structural parameters obtained from the refined model are presented in Table~\ref{tab:param}.
\begin{table}
\caption{\label{tab:param}Structural parameters of the hexagonal structure of CsC$_8$ at ambient conditions obtained from a Rietveld refinement. The corresponding stacking is A$\alpha$A$\beta$A$\gamma$. Thus, the $c$-axis lattice parameter corresponds to 3$I_c$.}
\begin{ruledtabular}
\begin{tabular}{cl}
Space group, No & \textit{P6$_2$22}, 180 \\
Lattice constants (\AA) & a= 4.9601, c=17.847 \\
Repeat distance (\AA) & I$_c$=c/3=5.949 \\
Atoms & Wyckoff positions \\
\hline
C & \textit{12k} (1/6, 1/3, 1/3)\\
C & \textit{6i} (5/6, 2/3, 0)\\
C & \textit{6i} (2/3, 1/3, 0)\\
Cs & \textit{3d} (1/2, 0, 1/2)\\
\end{tabular}
\end{ruledtabular}
\end{table} 

\subsection{High-pressure diffraction}

\subsubsection{Experimental procedure}

We have collected both high pressure X-ray and neutron powder diffraction data up to 14 GPa and 8.9 GPa respectively. Angle-dispersive X-ray diffraction (XRD) at the ID27 beamlime of the European Synchrotron Radiation Facility (ESRF) was performed using a monochromatic radiation ($\lambda$=0.26472 \AA) and a diamond anvil cell (DAC). The powder sample was loaded inside a 150 $\mu$m hole of a steel gasket with heavy mineral oil as pressure transmitting medium (PTM) and placed between the anvils of the DAC with diamond culets of 600 $\mu$m. Due to the high reactivity of the sample in the presence of oxygen and humidity, all loadings of the DAC were made inside an argon filled glove-box. The heavy mineral oil PTM ensures a further protection of the sample. A \textsc{ccd} detector was used to collect the high-pressure diffraction images. Calibration of the sample-detector distance and the tilt angle were done with a Si-diamond powder standard. The two-dimensional image files were azimutally integrated with the \textsc{fit2D} software yielding to one-dimensional intensity versus 2$\theta$~\cite{hammersley96}. The pressure was determined with the ruby fluorescence method~\cite{Mao78}. \textit{In situ} angle-dispersive neutron powder diffraction (NPD) experiments were carried out at the HRPT diffractometer~\cite{hrpt} located at the Swiss spalliation source (SINQ) using a VX5 Paris-Edinburgh Press~\cite{klotz04} in a radial configuration. In this geometry, the incident beam (monochromatic radiation of $\lambda$=1.494 \AA) and the diffracted beam pass through the TiZr gasket in the equatorial plane of the anvils. Neutron absorbing cubic boron-nitride (BN) anvils and null-scattering TiZr alloy gaskets were used in order to omit any signal from the pressure cell. Full details of the experimental setup are described in Ref.~\onlinecite{klotz05}. The assembly of the gasket and the loading of the press were completed inside an helium glove-box. Dried NaCl was used as pressure marker.

\subsubsection{X-ray diffraction results}
\begin{figure}[!t]
\includegraphics[trim= 0mm 0mm 0mm 0mm, clip, width=0.47\textwidth]{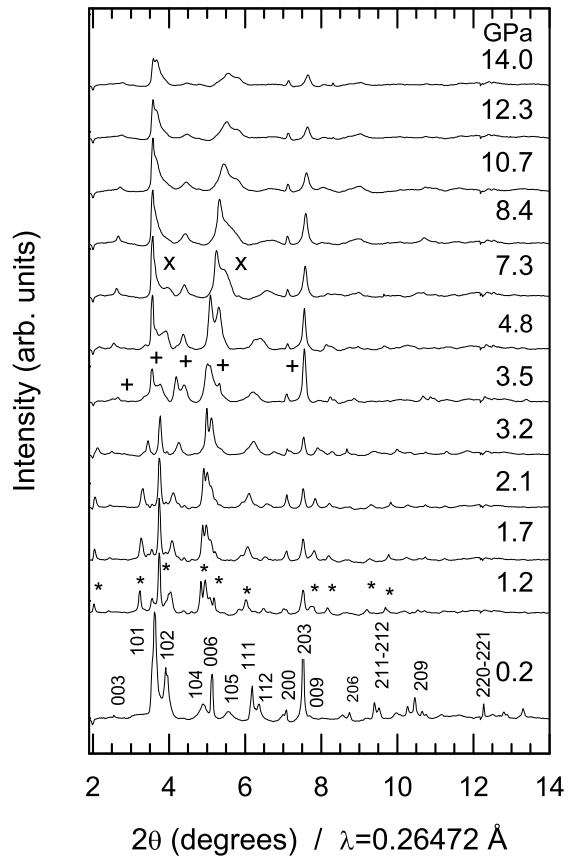}
\caption{\label{fig:xrd} Selected X-ray diffraction patterns of the CsC$_8$ compound as a function of pressure. Pressure in GPa is given at the right side of each curve. The first pressure pattern corresponds  to the hexagonal structure (CsC$_8$-I). The sample was confined in oil as pressure transmitting medium. Intensity was normalized for all patterns to the same counting time. Stars ($\star$), plus ($+$) and crosses ($\times$) symbols show the appearance of new Bragg peaks corresponding to CsC8-II, CsC8-III and CsC8-IV phases respectively (see Table~\ref{tab:resume}).}
\end{figure}

High pressure XRD of CsC$_8$ was performed up to 14.0 GPa. A sequence of selected diffraction patterns is displayed in Figure \ref{fig:xrd}. The first pattern recorded at 0.2~GPa matches the hexagonal ambient structure. No traces of oxide or hydroxide have been found. Inhomogeneous grain size distribution, preferential orientation effects and the identified stacking faults render a Rietveld refinement or even a Le Bail refinement extremely complex.  The very low intensity observed for the (003) and the (009) peaks makes difficult to detect their positions. Only the (006) line provides us a direct reliable information about the $c$-axis compressibility. Gradual transformations under increasing pressure is the most eye-catching feature as well as the overall decrease of intensity of the diffraction patterns. At 1.2~GPa, we observe the coexistence of the hexagonal phase with a new phase. This new phase is clearly established at 2.0 GPa. At 3.5 GPa, new changes in the diffraction pattern evidence a new phase transformation. Furthermore, we notice that the number of peaks of this last high pressure phase is lower than in the preceding phases, which could be associated to a higher cell symmetry or to a smaller lattice. As the pressure increases from 7.3~GPa to 14~GPa, the diffraction line broadening becomes more and more pronounced, and the profile becomes highly asymmetric.

\begin{figure}[!bt]
\includegraphics[width=0.45\textwidth]{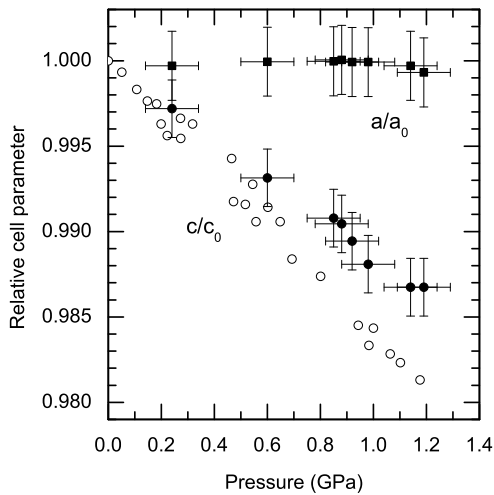}
\caption{\label{fig:compress} Compressibility of the hexagonal CsC$_8$ lattice parameters (solid symbols) derived from our X-ray diffraction experiment and of the $c$-axis reported by Wada (open symbols show two sets of data)~\cite{wada81}.} 
\end{figure}

Let us turn to the compressibility of the starting hexagonal phase at low pressures. Figure~\ref{fig:compress} shows the lattice parameters evolution with pressure as obtained from our XRD experiment.
The compressibility along the $c$-axis was found to be slightly lower than the one measured by Wada~\cite{wada81}. The difference must be moderated by considering the error bars and the use of different starting graphite samples for the two experiments. Wada used singe crystal graphite while we used powder graphite. We deduced a linear compressibility $\kappa_c$ of 1.079$\times$10$^{-2}$ GPa$^{-1}$ corresponding to a linear bulk modulus B$_c$ of 93~GPa (see table~\ref{tab:final}). The $a$ lattice parameter was found to be very rigid in this pressure range. We estimated the bulk modulus B$_v$=94~GPa, close to the linear inter-plane bulk modulus.
\begin{figure}[!b]
\begin{center}
\includegraphics[width=0.5\textwidth]{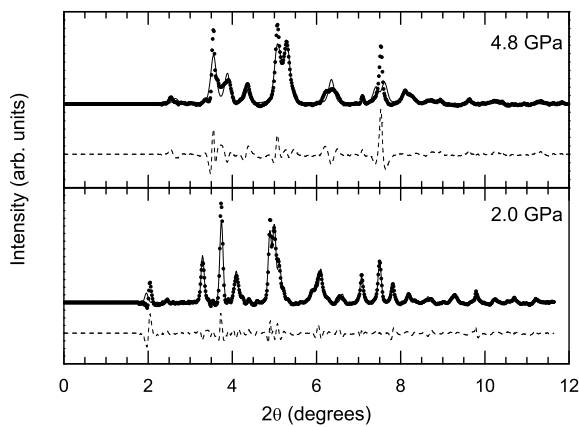} 
\end{center}
\caption{\label{fig:lebail} Le Bail refinements of the X-ray diffraction data for the high-pressure phases of CsC$_8$ at 2.0 GPa (bottom) and 4.8 GPa (top) corresponding to structural models shown in table \ref{tab:suma}.}
\end{figure}
\begin{table}
\caption{\label{tab:suma}Structural parameters of the possible candidate phases of CsC$_8$ appearing at 2-3 GPa and 4.8 GPa.}
\begin{ruledtabular}
\begin{tabular}{cdd}
 & \multicolumn{1}{c}{2.0~GPa} & \multicolumn{1}{c}{4.8~GPa} \\
\hline
Space group, No & \multicolumn{1}{c}{\textit{C222}, 21} & \multicolumn{1}{c}{\textit{Fddd}, 70}\\
Lattice constants (\AA) & \multicolumn{2}{c}{$\alpha=\beta=\gamma$= 90$^\circ$} \\
\hline
a & 5.24  & 4.39\\
b & 6.02  & 9.20   \\
c & 56.75 & 22.87    \\
\end{tabular}
\end{ruledtabular}
\end{table} 

Different attempts to determine the crystal structures of the high pressure phases appearing above 1.2 GPa and above 3.5 GPa were carried out. The presence of a diffraction line at very low angle $\sim$2$^{\circ}$ corresponding to a $d$-spacing of 7.69~\AA\ prompted us to look for a larger crystalline cell assuming a new stacking. We propose the space group \textit{C222} as a possible candidate for the first HP phase. This choice corresponds to a space group search procedure with the \textit{Crysfyre Suite}~\cite{crysfire} and \textit{Checkcell}~\cite{checkcell} softwares. Among the possible space groups obtained, \textit{C222} reproduced best the experimental data. The corresponding orthorhombic lattice is described in term of a supercell with parameters fitted within a Le Bail refinement (Table~\ref{tab:suma}). The second HP phase, observed between 3.5 and 8.0 GPa was found to be very close to the orthorhombic cell defined in RbC$_8$~\cite{lagrange78} and in CsC$_4$~\cite{nalimova96}, a superdense metastable phase of Cs GICs. We used the same space group \textit{Fddd} and Wyckoff positions as RbC$_8$ for the refinement model. This supposes a change of the stacking from $\alpha\beta\gamma$ to $\alpha\beta\gamma\delta$ with a corresponding $I_c$=c/4=5.72 \AA\ (a=4.39~\AA\, b=9.20~\AA\ and c=22.87~\AA) (Table~\ref{tab:suma}). Figure~\ref{fig:lebail} shows the Le Bail refinement and residuals for the two HP phases using the GSAS analysis software~\cite{toby01}. The two refinements are not completely satisfying: in the 4.8 GPa refinement, essentially broadening is not taken into account as we can see. And in the 2.0 GPa refinement, the position of the peak located around 2$^\circ$ is not well reproduced. In order to check the validity of our proposed lattices and to go deeper into the structural determination, i.e.\ to determine the atomic positions, better complementary and/or quality data are needed.
\begin{figure}[!t]
\includegraphics[width=0.47\textwidth]{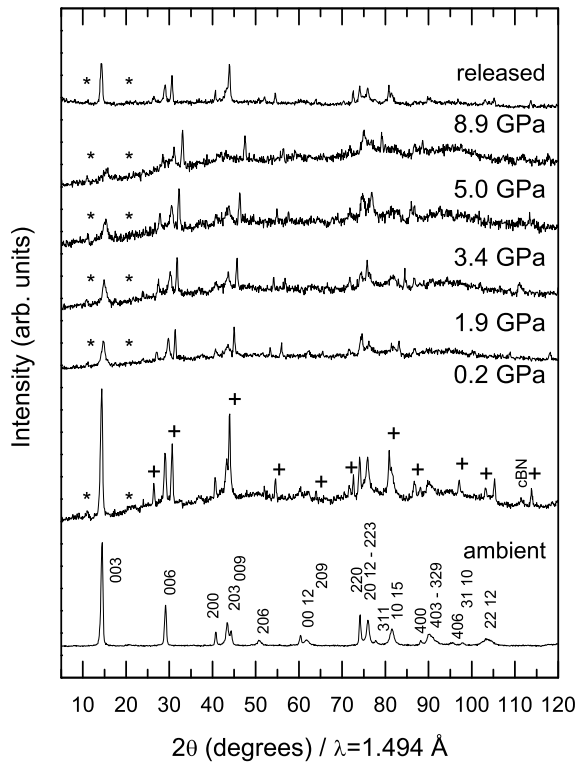} 
\caption{\label{fig:npdcsc8} Neutron powder diffraction patterns of CsC$_8$ as a function of pressure with NaCl ($+$) as pressure transmitting medium and pressure marker. The two stars ($\star$) indicate additional peaks (see text). The ambient diffraction pattern was taken in a vanadium container. The released pressure diffraction pattern was recorded in the press after the HP cycle.} 
\end{figure}
\begin{figure}[!b]
\begin{center}
\includegraphics[width=0.5\textwidth]{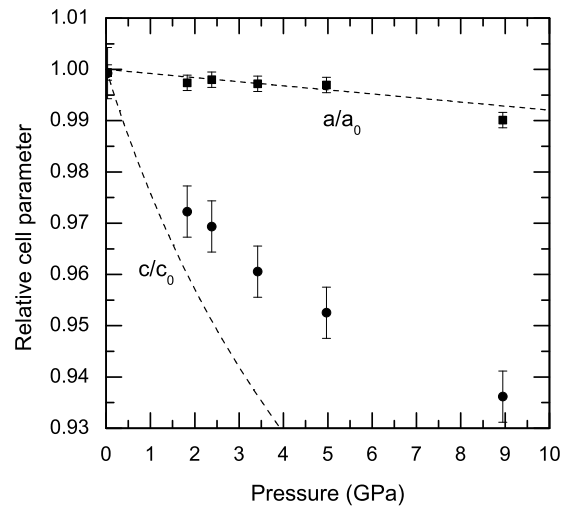} 
\end{center}
\caption{\label{fig:npd_fit}Evolution of the lattice parameters of CsC$_8$ (filled symbols) from NPD as a function of pressure compared to the graphite ones (dashed lines) from Ref.~\onlinecite{hanfli89}. The errors bars are provided from the Rietveld refinements.}
\end{figure}

\subsubsection{Neutron powder diffraction results}
We consider now the neutron diffraction data. In contrast to the X-ray case, where diffraction is dominated by the scattering cross-section of Cs, neutron diffraction is expected to provide a better information for the carbon sub-network because of the better balance between the scattering cross sections of carbon ($\sigma_{coh}$=5.551 barns) and Cs ($\sigma_{coh}$=3.690 barns). Due to the characteristics of the HP device, the maximal pressure was limited to 8.9 GPa. Three different experiments have been done: sample with NaCl (experiment no.1), sample with mineral oil (experiment no.2) and sample without any PTM (experiment no.3). The reversibility of the high pressure transformation was obtained in all the three cases after complete pressure release from 8.9 GPa as shown in Figure~\ref{fig:npdcsc8} for the first experiment. In the second experiment, increased incoherent background signal from the oil did not allow us to collect good quality data when increasing pressure. The pressure evolution of the $c$ and $a$ axis of CsC$_8$ from the first experiment is displayed in Figure~\ref{fig:npd_fit}. By using the pressure calibration curve of the first experiment as the pressure scale of the 3rd exp., its analysis leads to the same pressure dependence of the lattice parameters than for the 1st one. The compressibility of CsC$_8$ in the basal direction matches the compressibility of the graphite a-axis whereas the presence of intercalated Cs atoms produces a stiffening of the $c$-axis. In spite of the poor signal to noise ratio, appreciable changes in the intensity ratios between the (003) and (006) lines (see inset of Fig.~\ref{fig:npd_amb}) and in the broadening of the lines of the neutron data tend to support the observed transitions by XRD corresponding to the I$\rightarrow$II transition around 1~GPa and maybe III$\rightarrow$IV transition around 8 GPa.

Contrarily to the XRD experiments, no evident sign of phase transformation was found between 1.9 GPa and 8.9 GPa. We only noticed the presence of two very low-intensity peaks, around around 11$^{\circ}$ (7.8~\AA) and 22-24$^{\circ}$ (3.6-3.9~\AA) respectively  (see stars in Fig.~\ref{fig:npdcsc8}) respectively, when increasing pressure and which remain detectable after releasing the pressure to zero. The origin of these peaks is not clearly understood but no significant shift with pressure associated to these peaks is observed. Consequently, they cannot correspond to the (00$\ell$) lines of a stage-3/2 as it was previously reported in KC$_8$~\cite{fuerst83,kim86}. In addition, the extrapolated lattice parameters of the superdense CsC$_4$ phase (using the c- and a-axis compressibilities of Fig.~\ref{fig:npd_fit}) observed by Nalimova \textit{et al} around 0.5-1 GPa~\cite{nalimova96} cannot either explain these very weak peaks. Hence, they could result to the presence of a minority phase corresponding to an impurity or they could come from the HP environment.     

Combining X-ray and neutron diffraction is shown to be relevant to understand the behavior of CsC$_8$ under pressure. Differences between XRD and NPD results should be then attributed to two facts. First, XRD is essentially sensitive to the Cs layer order while NPD patterns are more dominated by the C atoms contribution. Secondly, the scattering geometries are different for the NPD (radial geometry) and XRD (axial geometry) experiments. It seems that due to the preferential orientation effect occurring with pressure in this lamellar compound, as it was observed for graphite~\cite{yagi92}, the radial geometry is better adapted to follow the (00$\ell$) lines.  At least two clear transitions have been evidenced at 2.0 GPa and 4.8 GPa by XRD associated to changes in the decoration of the Cs layer and in the stacking. On the other side, the NPD data show us that such transitions do not imply a structural change in the graphitic (host) sub-network (Fig.~\ref{fig:npdcsc8}). We observe a stiffening of the $c$-axis in CsC$_8$ in comparison with graphite. Hence, neutrons reveal that under high pressure, the initial average structure of CsC$_8$ initial phase is not strongly modified. 

\subsection{Raman spectroscopy}

The Raman spectra were collected with a LabRam HR800 spectrometer from Jobin-Yvon equipped with a \textsc{ccd} detector in the backscattering geometry. Spectra were excited using 632.8 nm radiation from an He-Ne air-cooled laser, Spectra Physics\texttrademark. Low power radiation from the laser source (few mW) was tuned to avoid de-intercalation effects caused by the local heating \cite{nemanich77,eklund77}. The laser beam was focused into the DAC through a Mitutoyo 50$\times$ objective down to a laser spotsize of approximatively 2 $\mu$m. The DAC was loaded inside a glove box to prevent the contamination of the highly sensitive sample. Stainless steel gaskets of 30 $\mu$m thickness were drilled to 135 $\mu$m hole diameter. Heavy mineral oil was used as a PTM in the pressure chamber. Ruby chips served for pressure calibration.

\begin{figure}
\subfigure[]{\label{fig:capia}\includegraphics[scale=0.4]{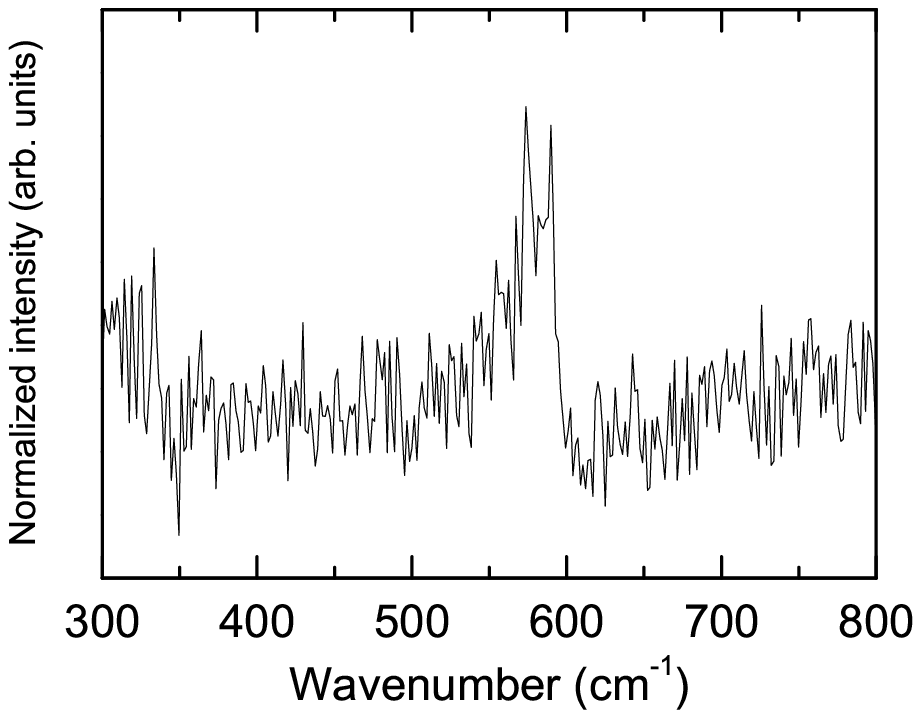}}
\subfigure[]{\label{fig:capib}\includegraphics[scale=0.4]{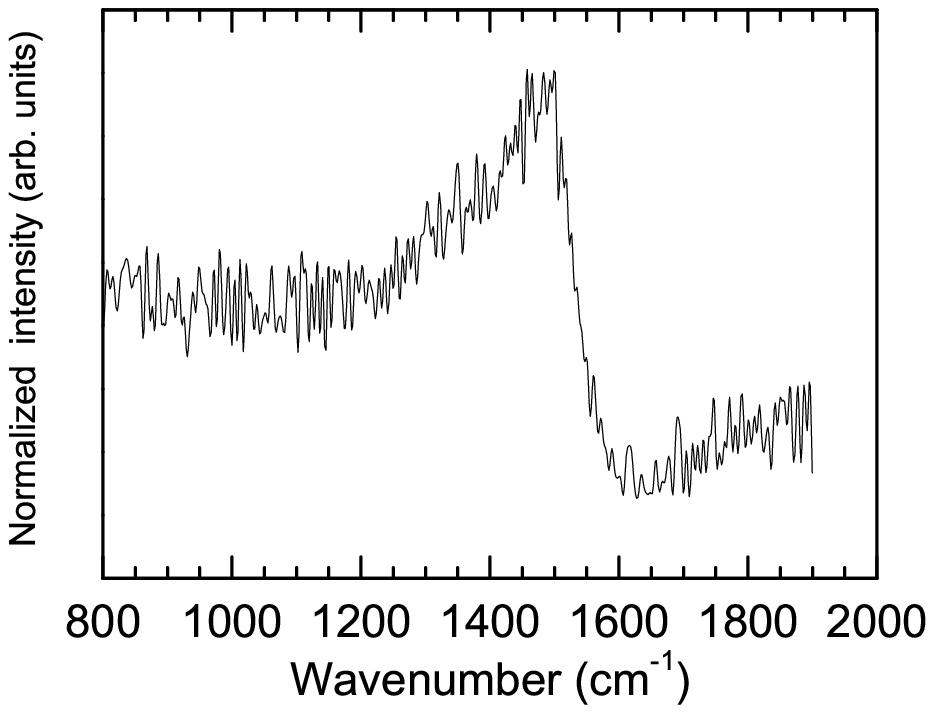}}
\caption{\label{fig:capi} High- (b) and low- (a) frequencies Raman signal of CsC$_8$ at ambient pressure and room temperature with a 632.8 nm laser excitation.}
\end{figure}

The CsC$_8$ Raman spectrum at ambient pressure is shown in Figure~\ref{fig:capi}. Characteristic features are a high-frequency broad structure (at approximatively 1500 cm$^{-1}$) and an low-frequency structure (near 560 cm$^{-1}$) in agreement with previous studies \cite{eklund77,nemanich77}. Intercalation of donor species like the alkali leads to a metallic behavior instead of the semi-metallic (graphite) character. This metallic character is more pronounced for rich alkali/C atoms compositions, typically for stage-1 compounds as in our case, making difficult the detection of the Raman scattering signal. 
Two first-order active Raman modes can be observed for pristine graphite, at 42~cm$^{-1}$ and at 1582~cm$^{-1}$. In GICs, the guest intercalate alkali layers act like a perturbation on the graphite modes thus leading to a shift in frequency \cite{Doll87} and turning on new active Raman modes. 
The striking high frequency broad asymmetric line is interpreted as a Breit-Wigner-Fano (BWF) resonance~\cite{eklund79}. This BWF line shape arises from an interaction of an electronic continuum with in-plane carbon-layer vibration corresponding to the $\Gamma$-point mode of pristine graphite. However, the origin of the singular structure in the vicinity of 560 cm$^{-1}$ which is in fact a triplet~\cite{Caswell79} rather than a doublet still remains unclear. This special feature was ascribed to a zone folding of M$_{1g}$ modes into the $\Gamma$-point for the ($2\times2$) superlattice corresponding to the MC$_8$ formula~\cite{eklund77} whereas the triplet was well accounted by disorder-induced scattering~\cite{Caswell79}.         

\begin{figure}
\includegraphics{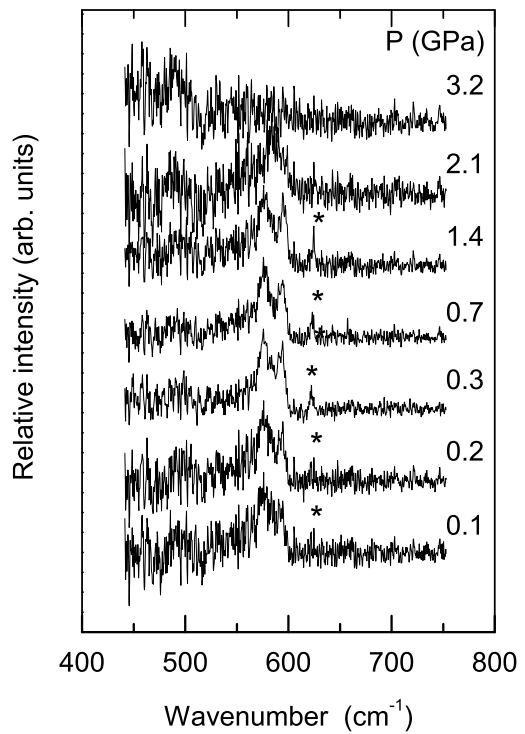}
\caption{\label{fig:raman2bf} Low-frequency Raman spectra of CsC$_8$ as a function of pressure. The last spectrum was collected after a HP cycle to 8.2 GPa. The star indicates the weak component consistent with the triplet observed by Caswell and Solin~\cite{Caswell79}.} 
\end{figure}

Figure~\ref{fig:raman2bf} shows the behavior of the low-frequency domain from 0.1 to 3.2 GPa. The first spectra collected in the DAC is in perfect agreement with the signal at ambient pressure reported in Figure~\ref{fig:capia}. The signal-to-noise ratio becomes of better quality as the pressure increases and we observe a better resolved triplet at 1.4 GPa. Then, the triplet gradually disappears above 1.4 GPa and cannot be detected beyond 3.2 GPa within our incertitude. 

\begin{figure}
\includegraphics{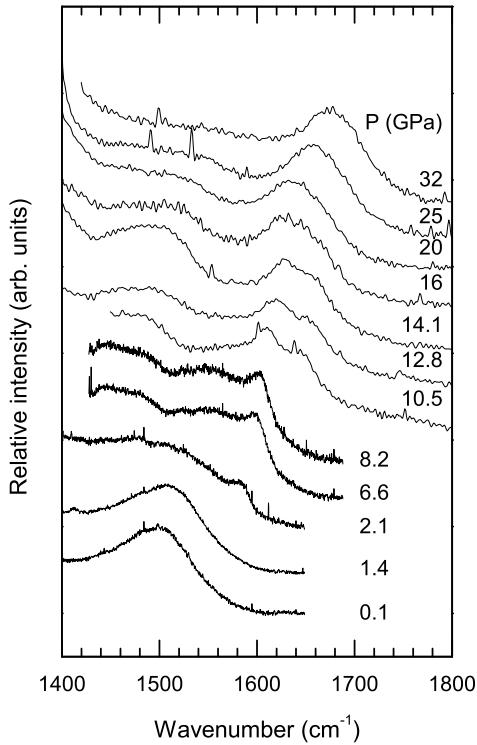}
\caption{\label{fig:raman1} High-pressure evolution of the high-frequency Raman spectra of CsC$_8$ at room temperature. From 0 to 8.2 GPa Raman spectra were collected using a 1800-line grating and a 600-line grating wase used from 10.5 to 32.0 GPa. The background is strengthened with Raman contribution of the oil.} 
\end{figure}

During the compression study, several changes of the high frequency mode have been observed. Selected high frequency Raman spectra of the CsC$_8$ compound taken at different pressures  are presented in Figure~\ref{fig:raman1} where the pronounced asymmetric broad shape associated to the BWF resonance is also shown. 
\begin{figure}
\includegraphics{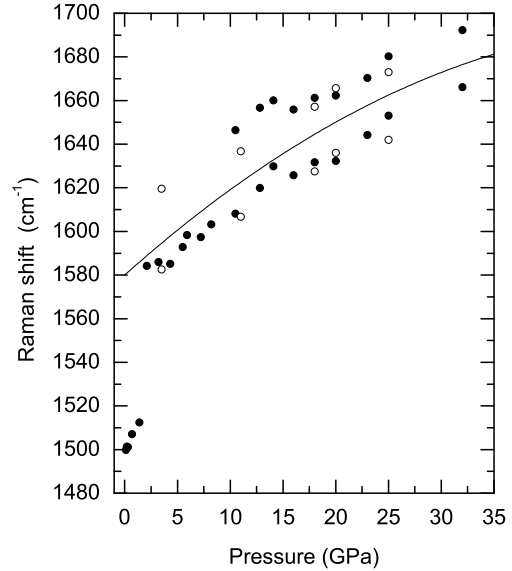}
\caption{\label{fig:modes} High-frequency Raman region peak positions of CsC$_8$ versus pressure. The straight line indicates the evolution of the $E_{2g}$ graphite mode with pressure from Ref.~\onlinecite{Schindler95}. Filled circles represent the compression whereas open circles show the decompression of the sample.}
\end{figure} 
In the low pressure region 1.4-2.1 GPa, the complete disappearance of the low-frequency triplet is accompanied by a large shift of the BWF resonance from around 1500 cm$^{-1}$ to 1590 cm$^{-1}$ (Fig.~\ref{fig:modes}). At 10.5 GPa, another peak appears close to the first one but slightly upshifted resulting in a doublet. Above 16 GPa and up to 32~GPa, it is extremely difficult to dissociate the two peaks, due to their large broadening. 

Upshifted Raman frequencies from the graphite mode (1582 cm$^{-1}$), visible above 10.5~GPa (Fig.~\ref{fig:raman1}), have been also observed in the case of diluted Cs-GICs. At ambient pressure, the stage-2 compound CsC$_{24}$ exhibits a Raman peak at 1598 cm$^{-1}$~\cite{nemanich77,eklund77}. In higher diluted compound like the CsC$_{36}$ stage-3, a doublet (1604-1579 cm$^{-1}$) is observed at ambient pressure~\cite{nemanich77,eklund77}. The lower frequency of this doublet is related to interior graphitic layers whereas the upper frequency is ascribed to the bounding graphitic layers. The high frequency doublet is systematically present for n$>$2 when the screening of the interior graphite layer by the bounding layer is not efficient anymore. 
\begin{figure}
\subfigure[]{\includegraphics[scale=0.4]{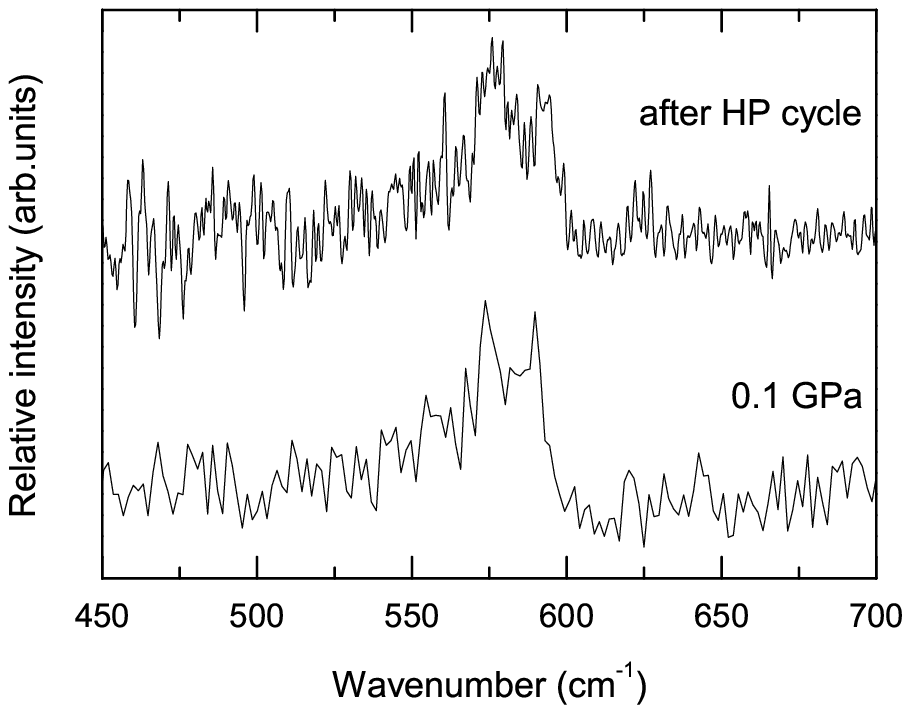}}
\subfigure[]{\includegraphics[scale=0.4]{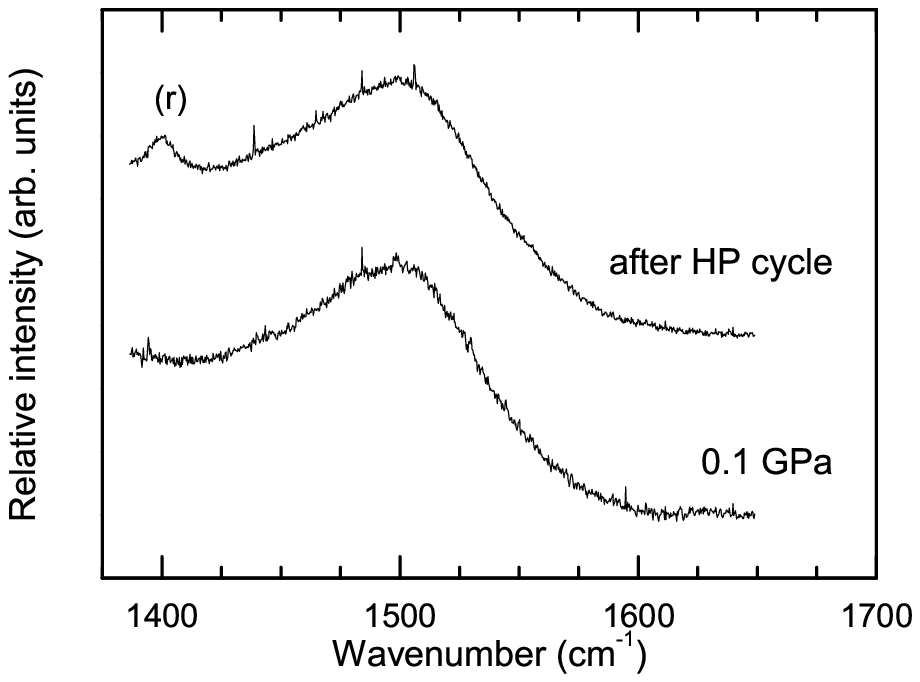}}
\caption{\label{fig:revers} Comparison of the low- (a) and high-frequencies (b) Raman spectrum of CsC$_8$ between the first pressure point at 0.1 GPa in the DAC and after the HP cycle to 8.2 GPa. Fluorescence of ruby is indicated by the (r) label.}
\end{figure}
The pressure behavior of the $E_{2g}$ graphite mode of graphite from Ref.~\onlinecite{Schindler95} is also shown in Fig.~\ref{fig:modes}. We remark that the peak appearing in the CsC$_8$ spectrum at 2.1 GPa around 1585~cm$^{-1}$ corresponds with the evolution of the graphite $E_{2g}$ peak and therefore could be associated to the interior graphitic layers mode observed in stage-$n$ with $n>2$. Consequently, the higher frequency of the doublet could be associated to the bounding layers mode.
\begin{figure*}[!t]
\includegraphics[width=0.8\textwidth]{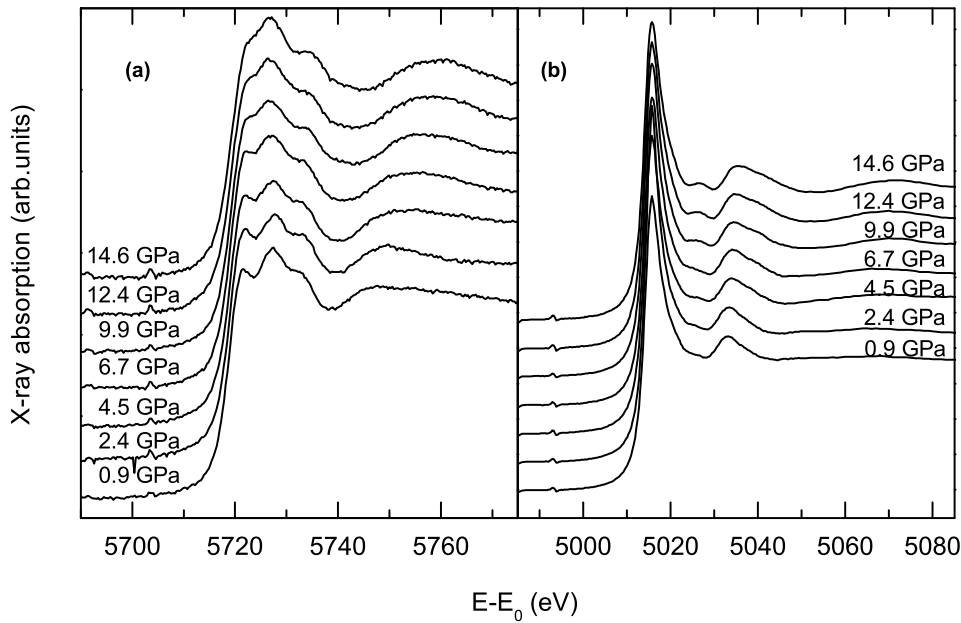}
\caption{\label{fig:xasL1L3} XANES evolution with pressure at both L$_1$-edge (a) and L$_3$-edge (b) of CsC$_8$ from our experiment at SLS.}
\end{figure*}
When the pressure is released from 32 GPa to 3.5 GPa, a reversible transformation accompanied with an hysteresis is observed (Fig.~\ref{fig:modes}). Due to technical reasons, we were not able to collect a Raman spectra at ambient pressure in the DAC. 
The initial Raman spectrum is recovered when the pressure is totally released from a maximum pressure of 8.2~GPa (Fig.~\ref{fig:revers}), indicating the reversible back-transformation to the ambient phase of CsC$_8$ as previously observed in our NPD experiments here reported. Thus, assuming de-intercalation process involving diffusion of the alkali atoms outside the graphitic host above 2 GPa, is incompatible with the observation of the CsC$_8$ low and high-frequencies features after the complete decompression. 

\subsection{X-ray absorption spectroscopy}
We carried out a X-ray absorption spectroscopy (XAS) experiment at the Cs L$_1$,L$_{2,3}$-edges on the microfocus Lucia beamline at SLS (Swiss Light Source, PSI, Villigen) under high pressure up to 14.5 GPa. Further detailed description of the beam-line set-up combined with high pressure are fully described in Ref.~\onlinecite{Flank06}. We worked in the energy range of 5-5.8 keV in the transmission configuration. DAC with stainless steel gaskets and hollowed diamonds were used to minimize X-ray absorption from the environment~\cite{itie05}. The sample was loaded with mineral heavy oil as PTM in a glove-box. Pressure was measured with the ruby fluorescence method. The three X-ray absorption Cs edges were  collected using the same set-up. In order to reduce the presence of glitches originated from the single crystal diamond anvils, the cell was properly oriented.

XAS is a powerful tool for the study of materials under very high pressure conditions~\cite{itie97}. In particular, it allows to obtain both electronic and structural informations 
of the local environment of atoms (in our case the Cs atoms). The Cs L$_{2,3}$-edges enable the study mainly of $d$-states whereas the L$_{1}$-edge $p$-final states are involved, as given by the dipolar selection rules. Information on guest-host hybridization or charge-transfer under pressure will be made available through their study. EXAFS oscillations were obtained for the L$_{3}$-edge, allowing, as it will be further discussed, to obtain qualitative information on the local structure evolution with pressure.

Figure~\ref{fig:xasL1L3} shows the normalized Cs L$_1$ and L$_3$ XANES evolution as function of pressure. The L$_2$-edge is not shown due to a lower signal-to-noise ratio and the similarity with the L$_3$-edge. For both Cs L$_1$- and L$_3$-edges, we observe a progressive evolution of XANES resonances with pressure. No sign of phase transition can be directly derived from XANES.
\begin{figure}
\includegraphics[scale=0.7]{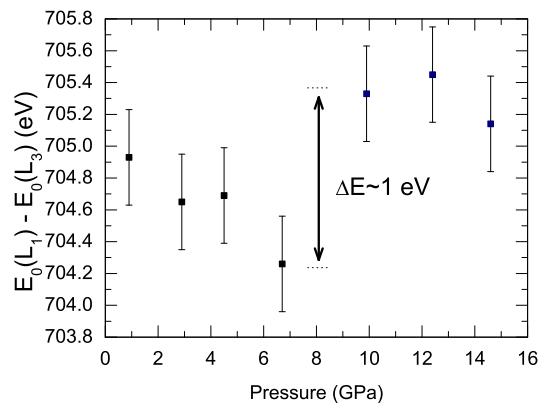}
\caption{\label{fig:seuilL13} Difference between the energy of the L$_1$- and L$_3$-Cs edges in CsC$_8$ as a function of pressure.}
\end{figure}
\begin{figure}[!b]
\includegraphics[scale=0.7]{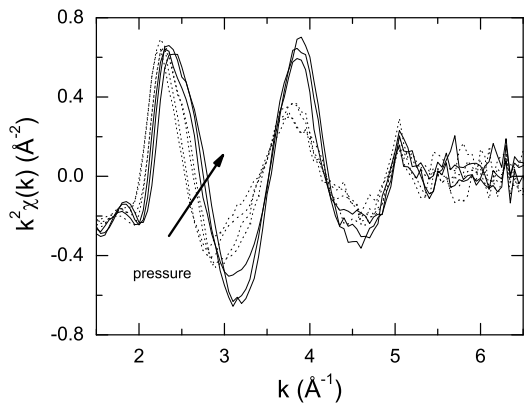}
\caption{\label{fig:exafsL3} Extracted Cs L$_3$-edge EXAFS oscillations as a function of pressure. The arrow shows increasing pressures from 0.9 to 14.6 GPa. Dotted lines corresponds to the low pressure phase and continuous lines to the high pressure phase (see text).}
\end{figure}
We followed the inflexion point for the different edges to monitor their energy position as a function of pressure. The absolute error bars in energy were too large to observe a significant evolution in the 0.9-14.5 GPa range. Nevertheless the accuracy of measurements was high enough to study the pressure dependence of the relative position between the L$_1$- and L$_3$-edges at each pressure point. In Fig.~\ref{fig:seuilL13} we plot the difference of energy position between the L$_1$- and L$_3$- edges as a function of pressure. An abrupt jump of approximatively 1 eV occurs between 7 and 10 GPa. This sudden change can be related to charge transfer effects between guest Cs atoms and the graphite host. Selection rules impose that this changes concern $p$-states and/or \textit{d}-states. However there is no further information allowing to establish the relative weight of charge transfer between the \textit{p}- and \textit{d}-channels.

The Cs L$_{3}$-edge EXAFS oscillations were extracted from the absorption spectra with the standard AUTOBK procedure~\cite{Newville93}. In Figure~\ref{fig:exafsL3} are shown the obtained EXAFS oscillations which are expressed in terms of the photoelectron wave number $k=1/({\hbar}\sqrt{2m(E-E_{0}}))$ where $\textit{m}$ is the electron mass and $\textit{E$_{0}$}$ is the absoprtion edge position. The limited extent of the signal originates from dynamic or static disorder as well as in the low-Z values of the main photoelectron scatterers (C atoms). In spite of this, we can appreciate important changes in the amplitude of EXAFS oscillations in the region between 2.5 and 4.5 \AA$^{-1}$. These changes correspond to a sudden increase of the amplitude of the signal that takes place between 7~GPa and 10~GPa and that can be associated with the changes in the relative position of the absorption edges (Fig.~\ref{fig:seuilL13}).
\begin{figure}[!h]
\includegraphics[width=0.4\textwidth]{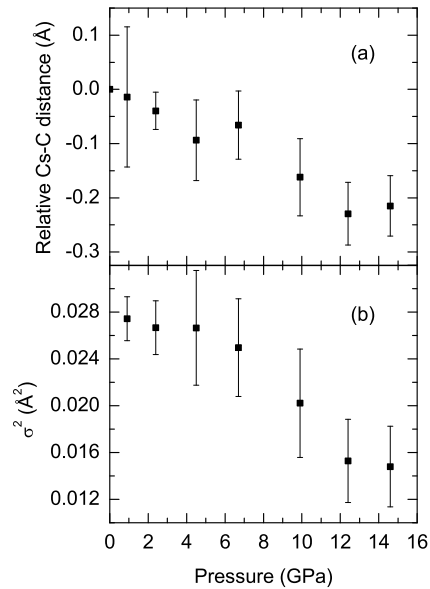}
\caption{\label{fig:xas_csc8} Pressure-evolution of the relative Cs-C first neighbors distance (a) and the pseudo-Debye Waller $\sigma^2$ (b). The EXAFS signal was $k^2$ weighted and Fourier transformed [2.5, 6.0] \AA$^{-1}$ to real space and finally fitted in the range [1, 3] \AA.} 
\end{figure}
\begingroup
\squeezetable
\begin{table*}
\caption{\label{tab:resume}Summary of the principal changes observed in the CsC$_8$ compound as a function of pressure.}
\begin{ruledtabular}
\begin{tabular}{@{\hspace{5mm}}c@{\hspace{8mm}}|c@{\hspace{8mm}}|p{2.5cm}|p{7.5cm}}
 & P-stability (GPa) & Observed structures & Most significative changes \\
\hline
CsC$_8$-I & 0.0-1.2 & \textit{P6$_2$22} & Linear compressibility   \\ 
  &        &    stage-1 ($\alpha\beta\gamma$)&          \\ \hline
CsC$_8$-II & 1.2-3.5 & \textit{C222}\footnotemark[1] &  Change in the Cs decoration accompanied with small displacements and strong modification of the Raman signal.     \\ 
          &        & stage-1 + ?    &  Abrupt change in the relative intensity of (003) and (006), strong broadening of the lines in NPD. \\ \hline    
CsC$_8$-III & 3.5-8.0 & \textit{Fddd}\footnotemark[1]  & Changes in the Cs decoration (slight displacements), in the stacking ($\alpha\beta\gamma\delta$) and possible intermediate structure between II-III.     \\ 
            &          & stage-1 + ? & Broadening of the diffraction lines in NPD. \\ \hline          
CsC$_8$-IV & 8.0-32.0 & stage-$n$ ($n>$2) & Electronic changes observed by EXAFS and Raman at the transition. \\
  &        &  + other phase? & Significant broadening of the diffraction lines and very low intensity signal in XRD. No evidence of the presence of the stage-1. Doublet observed in the Raman signal.\\
\end{tabular}
\end{ruledtabular}
\footnotetext[1]{Proposed structures}
\end{table*}
\endgroup

Due to the small $k$-range window for the EXAFS oscillations and the presence of disorder in the material, we have restricted the quantitative analysis to the first shell of carbon neighbors around the absorbing Cs atoms. We used the \textit{ab initio} \textsc{feff 8.0} code~\cite{Ankudinov98} (based on Hedin-Lundqvist exchange potential) to generate the scattering functions (amplitude and phase shift of the photoabsorbed and backscattered atoms). The EXAFS oscillations corresponding to the first shell Cs-C was Fourier filtered and fitted with the first single diffusion path using the \textsc{feffit} program~\cite{Newville95}. In addition to the Cs-C first neighbors distance (Fig.~\ref{fig:xas_csc8}(a)) and the associated pseudo Debye-Waller factor, we considered also the third cumulant as the only free parameters in our fitted EXAFS expression. We used the cumulant expression implemented in the \textsc{feffit} program based on an anharmonic effective pair potential. The observed change in magnitude of the EXAFS oscillations translates into a modification of the pseudo Debye-Waller contribution as shown in Fig.~\ref{fig:xas_csc8}(b). Nevertheless, adding local disorder effect (thermal or structural) with the $C_3$ cumulant allows to reproduce the graphite inter-plane distance in a good agreement with our other data from XRD and NPD (Fig.~\ref{fig:inter_final}). 

\section{Discussion}
As it is shown in Table~\ref{tab:resume}, we can distinguish four domains of stability characterized by modifications in the crystalline structure (as determined by XRD, NPD or by Raman spectroscopy) or by modifications of the electronic structure (evidenced either by XAS or Raman spectroscopy) of CsC$_8$. The most remarkable facts are the smooth evolution both of the $c$ lattice parameters obtained by NPD up to 9~GPa and of the Cs-C distances given by EXAFS up to 14.6 GPa even if the later case the incertitudes remain very large. The high sensitivity of XRD to the Cs atoms ordering with respect to NPD let us conclude that the two first phase transitions up to 14.0 GPa are related to slight modifications of the arrangement of Cs atoms, which involve the modification of the correlation between the different layers. In fact, the best fits to the XRD data at 2.0 GPa was performed with a very large unit cell (a=5.24~\AA\, b=6.02~\AA\ and c=56.75~\AA, SG \textit{C222}, No 21) which in addition is not commensurate with the graphitic sub-network.

The I$\rightarrow$II phase transition is particularly clear as it is well defined and both observed by XRD and by Raman spectroscopy. The confrontation of XRD and NPD data points out a modification of Cs decoration without changes in the graphitic plane stacking, because in spite of their relative intensity changes, (00$\ell$) reflections remain under the whole compression. The important change in the high-frequency Raman peak position at the I$\rightarrow$II phase transition could be interpreted as a stage modification. Nevertheless, this appears as contradicting with the NPD data. A possible interpretation is that Raman spectroscopy being more surface sensitive in this metallic compound, staging transitions could be observable at the grain surface. The II$\rightarrow$III phase transition, as was already discussed, has only been detected through XRD and it may be accompanied by the presence of an intermediate phase. As for phase II, phase III seems to involve a very large crystalline lattice as suggested by the appearance of low-angle peaks. NPD does not show any difference between phase I, II and III, so again CsC$_8$-III is most probably associated to changes in the decoration of the Cs sub-network. The last phase transition, III$\rightarrow$IV around 8 GPa, is related to changes of the electronic structure, that could imply a stage transition. It is clearly observed by XAS with the associated 1~eV shift between the Cs L$_2$ and L$_3$ absorption edges, and by Raman spectroscopy with an interior graphitic layers mode and a bounding layers mode, characteristic of a high-stage $n>$2. This transition occurs in all the sample since XAS is a bulk probe. We shall again mention that the observation of a doublet in the Raman signal from around 10~GPa involves the grain surface of the metallic sample. Furthermore, we cannot exclude that the doublet result from the local laser heating. As we do not have NPD data above 9 GPa and fits of the XRD data in this region are not reliable, we cannot assure the persistence of the stage-1. Other structural changes associated to this last transition cannot be excluded. Indeed, $n\rightarrow n$+1 staging transition implies the presence of an additional phase to maintain the density of Cs atoms.
\begin{table}[!b]
\caption{\label{tab:final}Axial compressibility and bulk modulus of CsC$_8$. Incertitudes for our values are of 10\%.}
\begin{center}
\begin{ruledtabular}
\begin{tabular}{lccccr}
Techniques & B$_c$ (GPa) & $B_c'$ & B$_v$ (GPa) & $B_v'$ & P-range \\
\hline
XRD this work\footnotemark[1]& 93& 0  &  94 & 0& 0-1.1 GPa\\
INS\footnotemark[2]  & 58 &  & &  &  ambient   \\ 
XRD\footnotemark[3]  & 64 & & &  & 0-1.1 GPa \\
NPD this work & 72 & 8 & 70 & 6 & 0-8.9 GPa\\
\end{tabular}
\end{ruledtabular}
\end{center}
\footnotetext[1]{Linear fit.}
\footnotetext[2]{Inelastic Neutron Scattering (INS) from reference~\cite{zabel82}.}
\footnotetext[3]{Reference~\cite{wada81}.}
\end{table} 

Let us now turn to the compressibility of the structure. Compressibility of phase I, characterized either by the volume bulk modulus or by the linear bulk modulus of the $c$-axis is given in Table~\ref{tab:final}. Our values are compared with other values from the literature which propose higher compressibilities. As mentioned, the neutron diffraction data (Fig.~\ref{fig:npdcsc8}) show a smooth variation of the graphitic-like cell parameters with pressure that do not appear to be affected by any clear phase transitions within the calculated incertitude. With a Murnaghan fit up to the highest pressure measured in the NPD experiment, i.e.\ 8.9 GPa, we obtain a bulk modulus equal to 72 GPa, which, as can be seen in Table~\ref{tab:final}, is closer to the value published by Wada~\cite{wada81}.
\begin{figure}[!t]
\begin{center}
\includegraphics[width=0.4\textwidth]{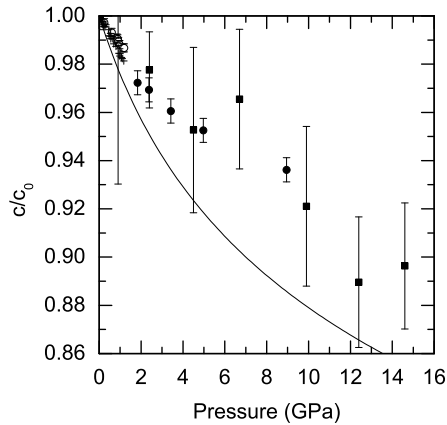}
\end{center}
\caption{\label{fig:inter_final}Dependence of the CsC$_8$ graphitic inter-plane distance with pressure combining both diffraction (XRD and NPD) datas indicated by empty and solid circles and local spectroscopy (EXAFS) datas indicated by solid squares. Crosses from 0 to 1.1~GPa correspond to the data of Wada~\cite{wada81}. The compressibility of graphite (solid line) from Ref.~\onlinecite{hanfli89} is plotted for comparison.}
\end{figure}


In addition, we deduced the graphitic inter-plane compressibility from the projection of the Cs-C distance on the $c$ axis by EXAFS. Taking into account an anharmonic correction through the third cumulant related to this distance allows to provide a reasonable agreement with the compressibility obtained by the diffraction experiments within the incertitudes (Fig.~\ref{fig:inter_final}). Introduction of disorder variations in the EXAFS expression to fit as good as possible the diffraction data is a consequence of the changes in the Cs atoms decoration observed in the XRD patterns. We observe in Figure~\ref{fig:xas_csc8}(b) from the pseudo Debye-Waller factor evolution that compression leads to a progressive reduction of the static or/and dynamic disorder of the local Cs structure.

\section{Concluding remarks}

Combining different techniques, we have studied the pressure behavior of the guest and host sub-networks and the guest-host interaction of the CsC$_8$ intercalated graphite in the 0-32 GPa range. Three pressure-induced transformations have been clearly observed. The high reactive nature of the sample coupled with its ill order nature, made the use of complementary experimental techniques a necessity. The combination of these techniques, in which different selectivities apply, has been essential for the understanding  of the observed transitions. We have given evidences of evolutions in both the electronic structure (using Raman and XAS spectroscopies) and in the crystallographic structure (XRD and NPD) occurring in the studied pressure range. The main observed facts can be summarized as follow: (i) anisotropic compressibility, (ii) stiffening of the $c$-axis in CsC$_8$ in comparison to graphite up to 8.9 GPa due to the presence of intercalated Cs atoms, (iii) subtle re-arrangements of the Cs atoms layer lead to a complex lattice, (iv) a re-organization of the stacking probably occurs (from hexagonal $\alpha\beta\gamma$ to orthorhombic $\alpha\beta\gamma\delta$ at 4.8~GPa for example) and (v) hybridization between Cs atoms and graphite layers could explain the charge transfer observed in XAS around 8~GPa. Finally, with exception of the appearance of an impurity phase, it seems that these transformations are reversible at least up to $\sim$8 GPa, indicating that they do not result from a de-intercalation process.

\begin{acknowledgments}
This work was supported by the French Minist\`ere D\'el\'egu\'e à la Recherche et aux Nouvelles Technologies
and the CNRS through grants "ACI-2003 n. NR0122" and "ACI-2003 n. JC2077". We thank gratefully H. Feret (LPMCN, France) for his assistance and technical support. This research project has been supported by the European Commission under the 6th Framework Programme through the Key Action: Strengthening the European Research Area, Research Infrastructures. Contract no: RII3-CT-2004-506008. This work is based on experiments performed at the Swiss spallation neutron source SINQ,  Paul Scherrer Institute, Villigen, Switzerland.
\end{acknowledgments}

\bibliography{article}

\end{document}